\documentclass[conference]{IEEEtran}
\IEEEoverridecommandlockouts
\usepackage{cite}
\usepackage{amsmath,amssymb,amsfonts}
\usepackage{algorithmic}
\usepackage{graphicx}
\usepackage{textcomp}
\usepackage{xcolor}
\def\BibTeX{{\rm B\kern-.05em{\sc i\kern-.025em b}\kern-.08em
    T\kern-.1667em\lower.7ex\hbox{E}\kern-.125emX}}

\newcommand{\figintroJSCC}{
  \begin{figure}
    \centering
    \includegraphics[width=1\columnwidth]{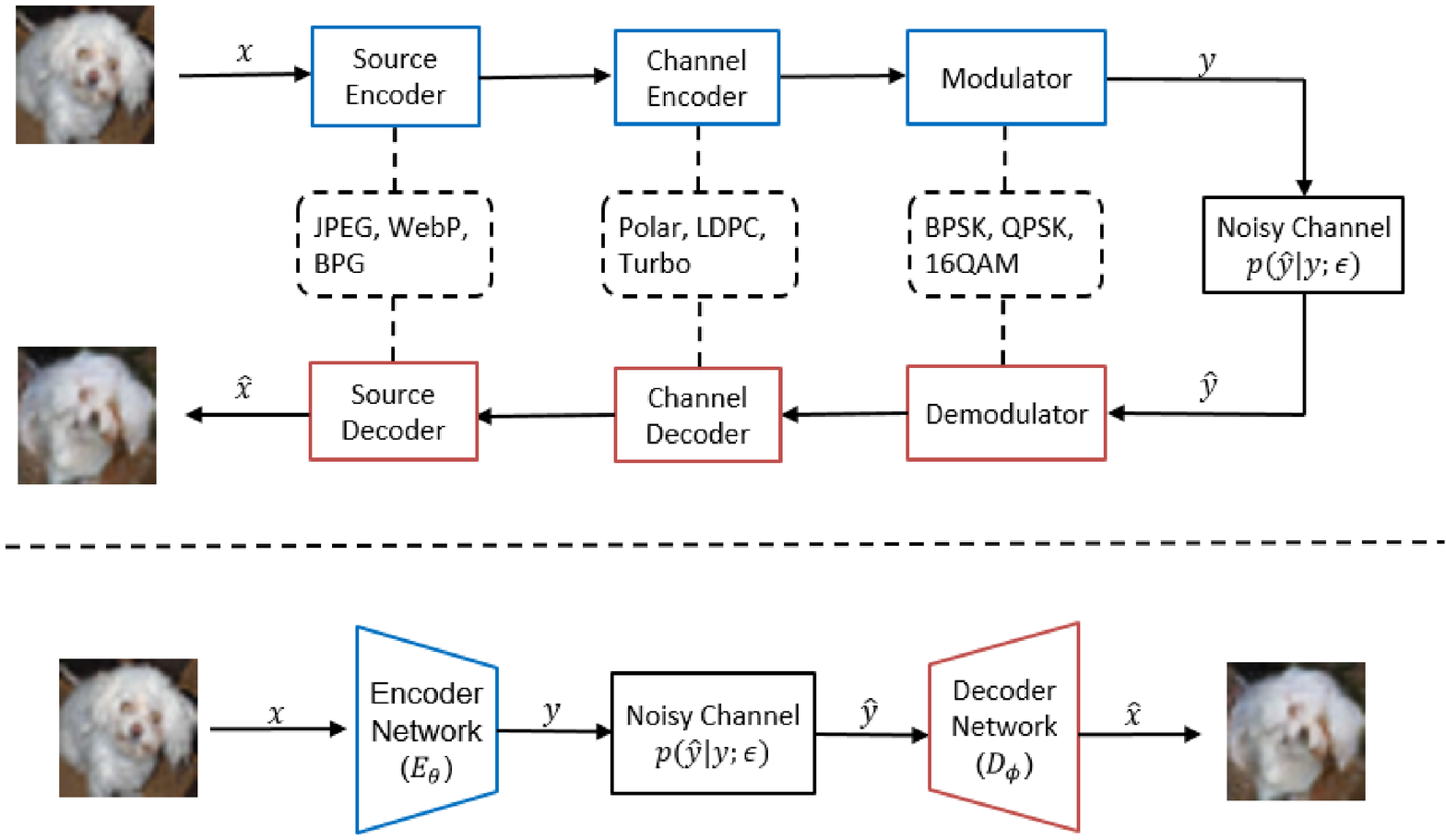}
    \caption{Block diagrams of wireless image transmission schemes. Top: Traditional separate source and channel coding scheme. Bottom: Deep learning based joint source channel coding (JSCC) scheme.}
    \label{fig:fig_intro}
  \end{figure}
}

\newcommand{\figofdmlarge}{
  \begin{figure*}[t]
    \centering
    \includegraphics[width=\linewidth]{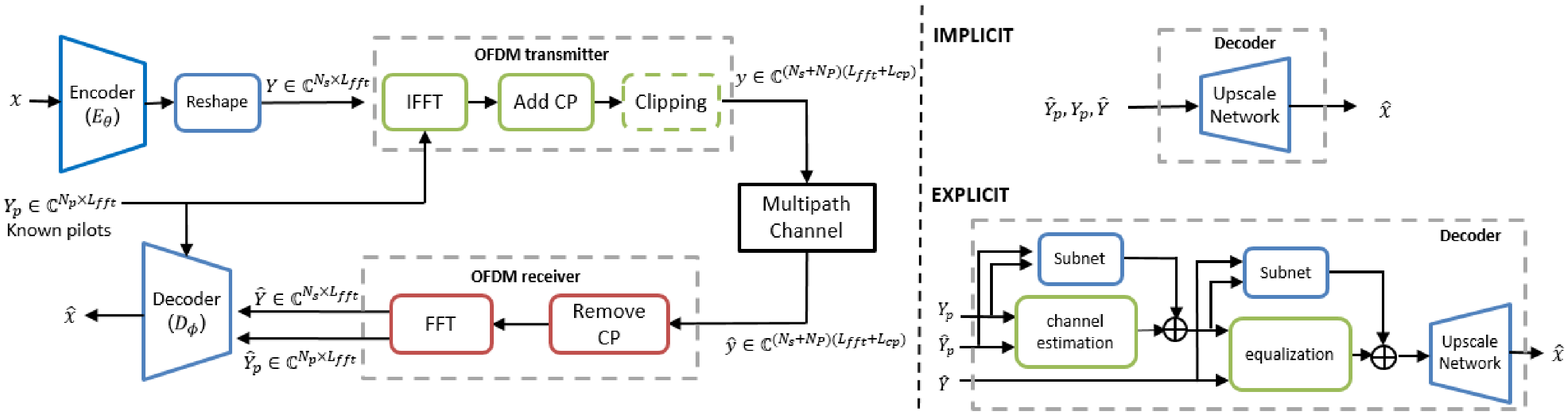}
    \vspace{-0.6cm}
    \caption{Left: Flow diagram for the proposed JSCC framework with OFDM extension. Right: Two decoder structures considered in this paper. IMPLICIT: directly concatenate everything and feed them to the decoder network. EXPLICIT: introduce explicit channel estimation and equalization methods with additional residual connections. }
    \label{fig:fig_ofdm}
  \end{figure*}
}

\newcommand{\fignet}{
  \begin{figure}
    \centering
    \includegraphics[width=1\columnwidth]{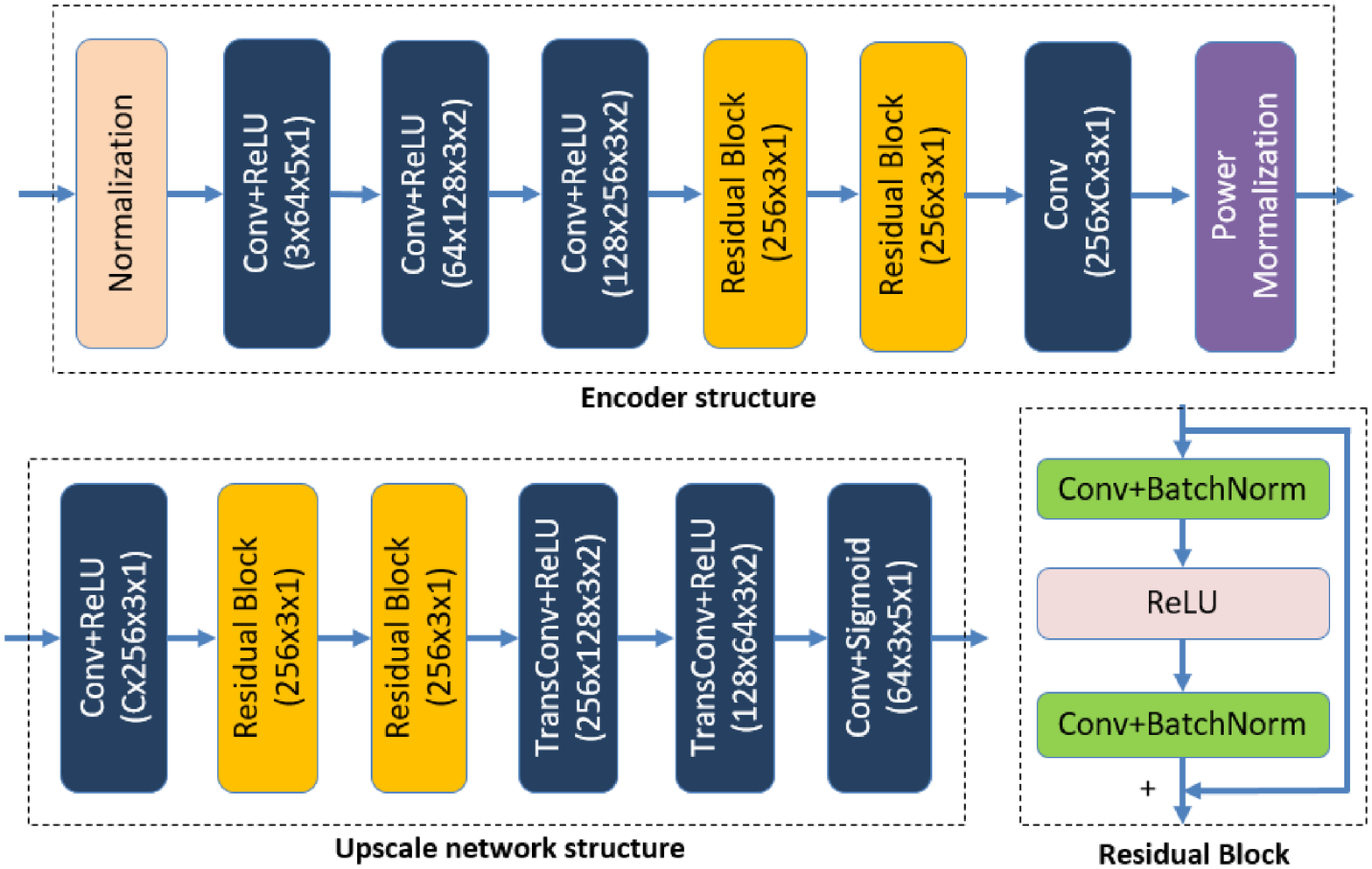}
    \vspace{-0.6cm}
    \caption{Network structure for the proposed method. There is a batch normalization layer between each convolutional layer and ReLU activation function (although not shown in the figure).}
    \label{fig:fig_net}
  \end{figure}
}

\newcommand{\figlr}{
  \begin{figure}[t]
    \centering
    \includegraphics[width=0.8\columnwidth]{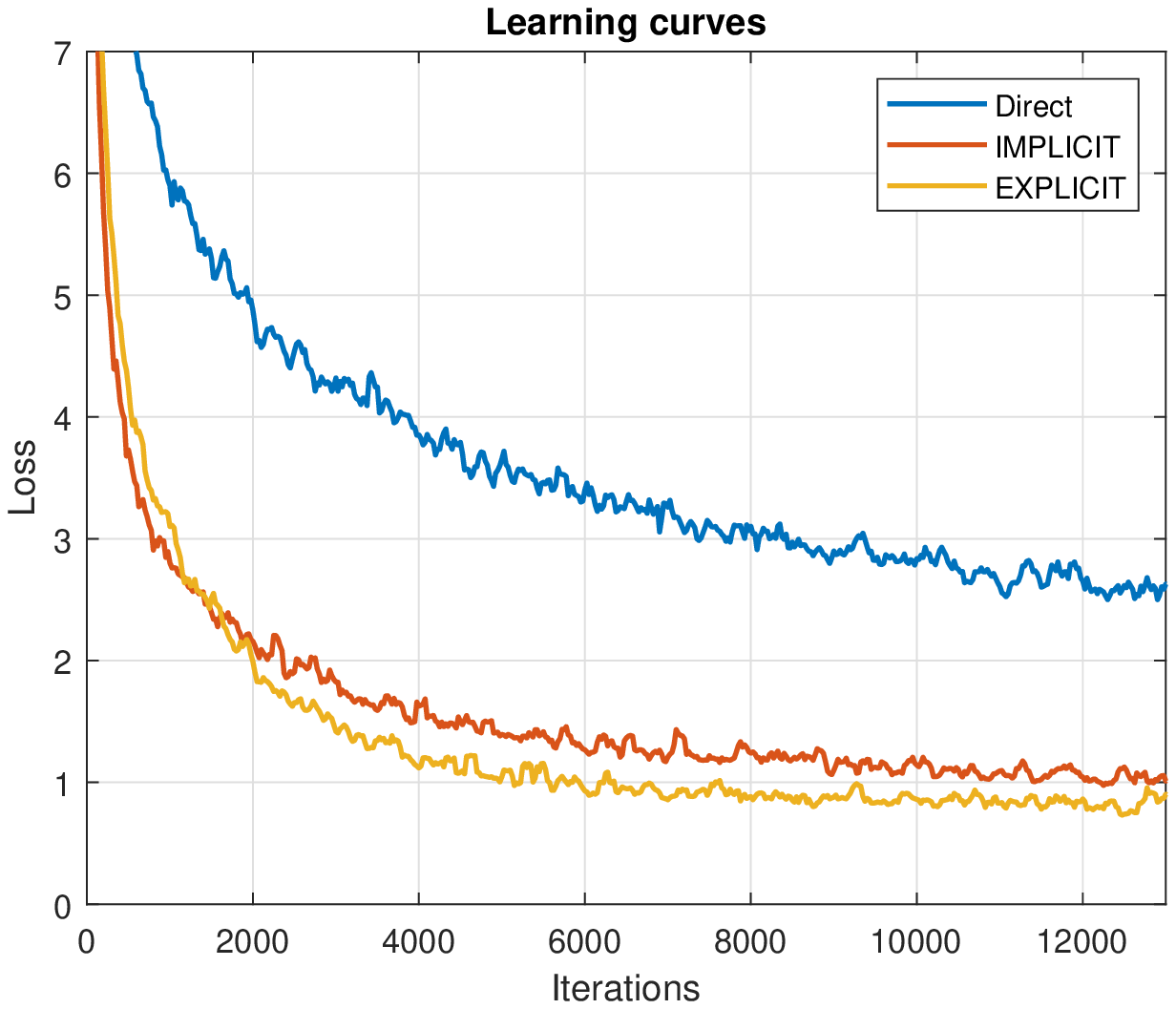}
    \vspace{-0.3cm}
    \caption{Learning curves for three methods with a SNR of $20$dB and $N_s=6$ for CelebA dataset.}
    \label{fig:fig_lrcurve}
  \end{figure}
}

\newcommand{\figlowpsnr}{
  \begin{figure}[t]
    \begin{minipage}[b]{.48\linewidth}
      \centering
      \centerline{\includegraphics[width=4.7cm]{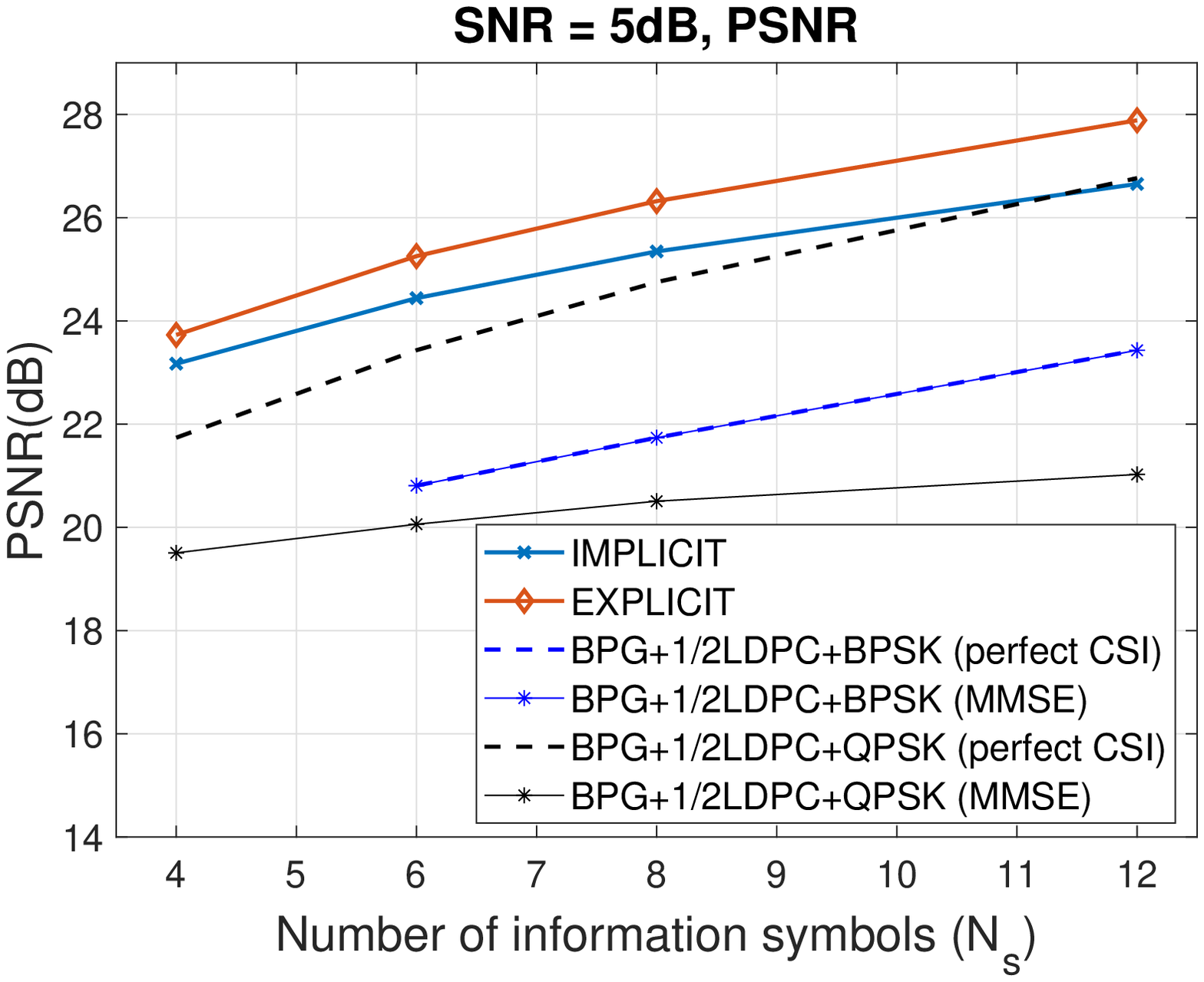}}
    \end{minipage}
  \begin{minipage}[b]{.48\linewidth}
    \centering
    \centerline{\includegraphics[width=4.7cm]{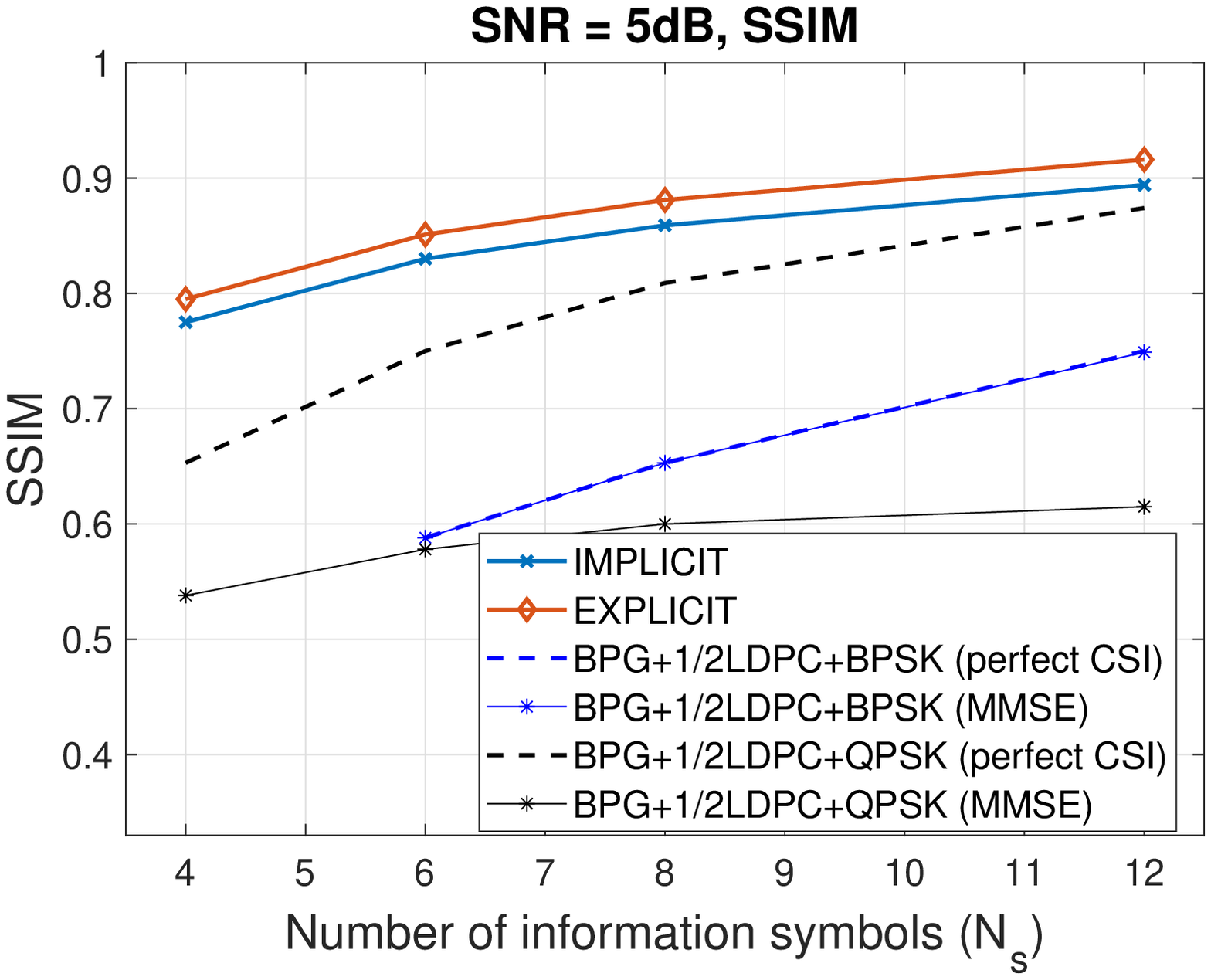}}
  \end{minipage}
\caption{Performance of the proposed method on CIFAR-10 with respect to the number of information symbols ($N_s$) for a fixed SNR of 5dB}
\label{fig:fig_lowsnr}
\end{figure}
}

\newcommand{\fighighpsnr}{
  \begin{figure}[t]
    \begin{minipage}[b]{.48\linewidth}
      \centering
      \centerline{\includegraphics[width=4.7cm]{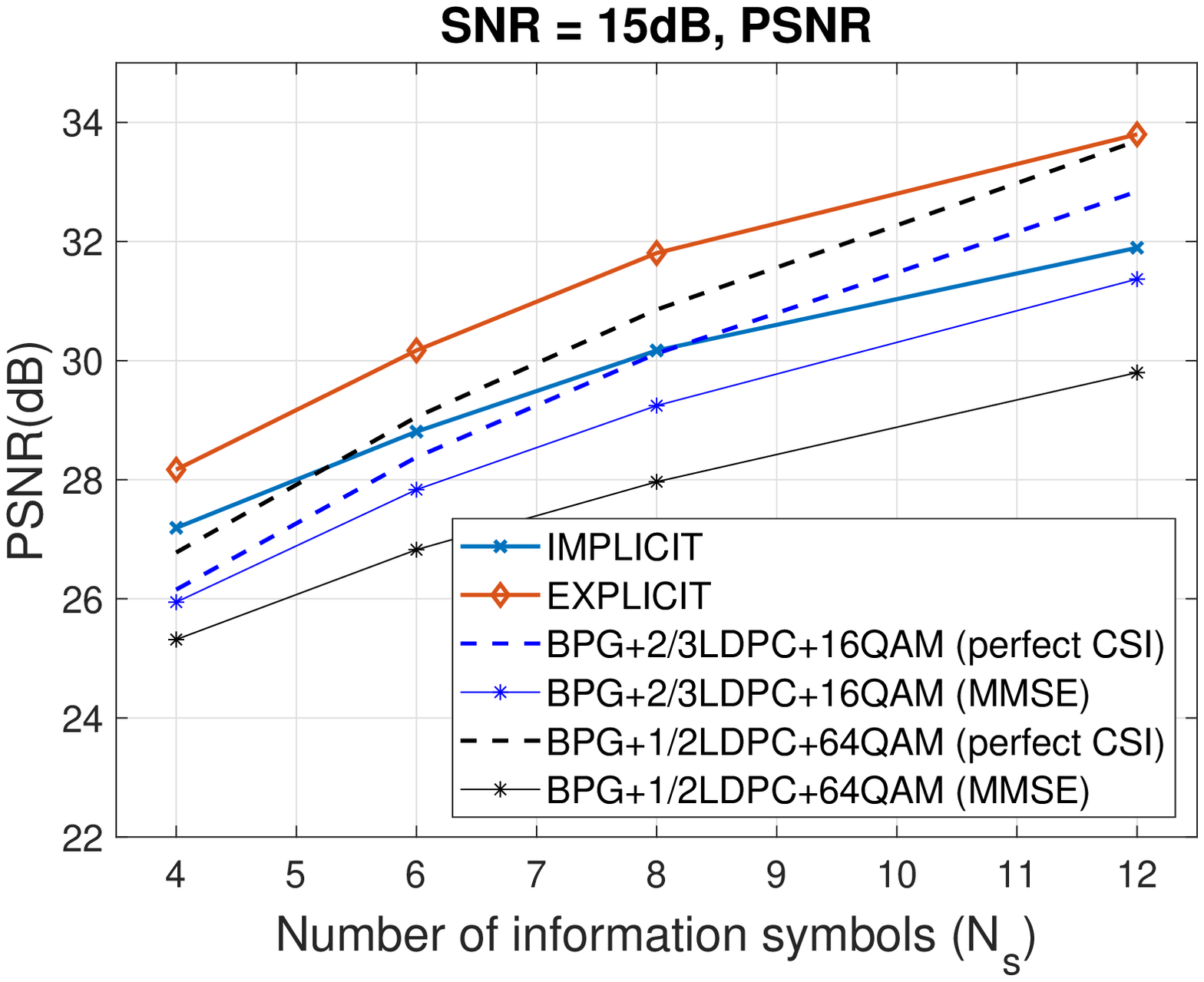}}
    \end{minipage}
  \begin{minipage}[b]{.48\linewidth}
    \centering
    \centerline{\includegraphics[width=4.7cm]{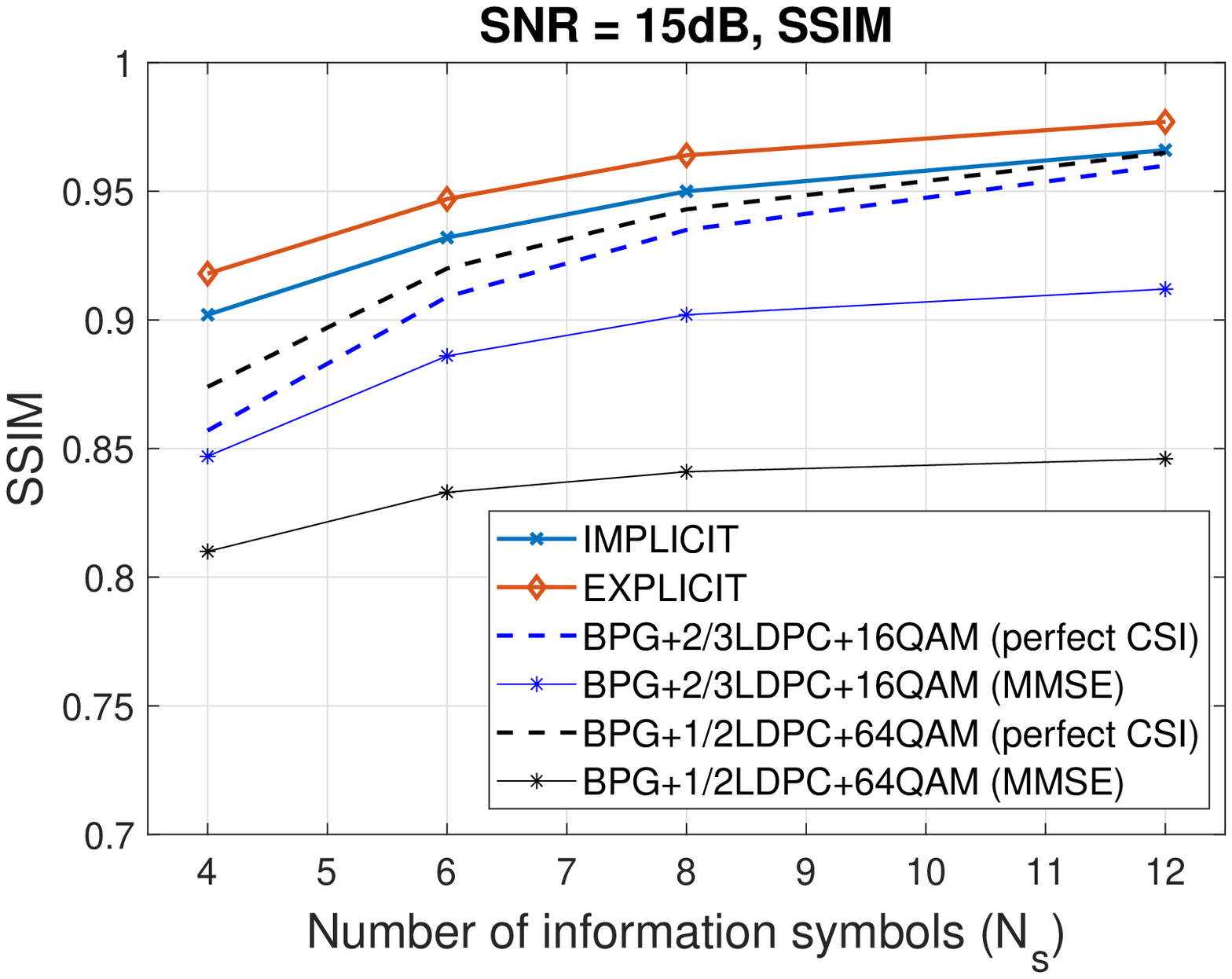}}
  \end{minipage}
  \hfill
\caption{Performance of the proposed method on CIFAR-10 with respect to the number of information symbols ($N_s$) for a fixed SNR of 15dB}
\label{fig:fig_highsnr}
\end{figure}
}

\newcommand{\figpsnrfr}{
  \begin{figure}[t]
    \begin{minipage}[b]{.48\linewidth}
      \centering
      \centerline{\includegraphics[width=4.7cm]{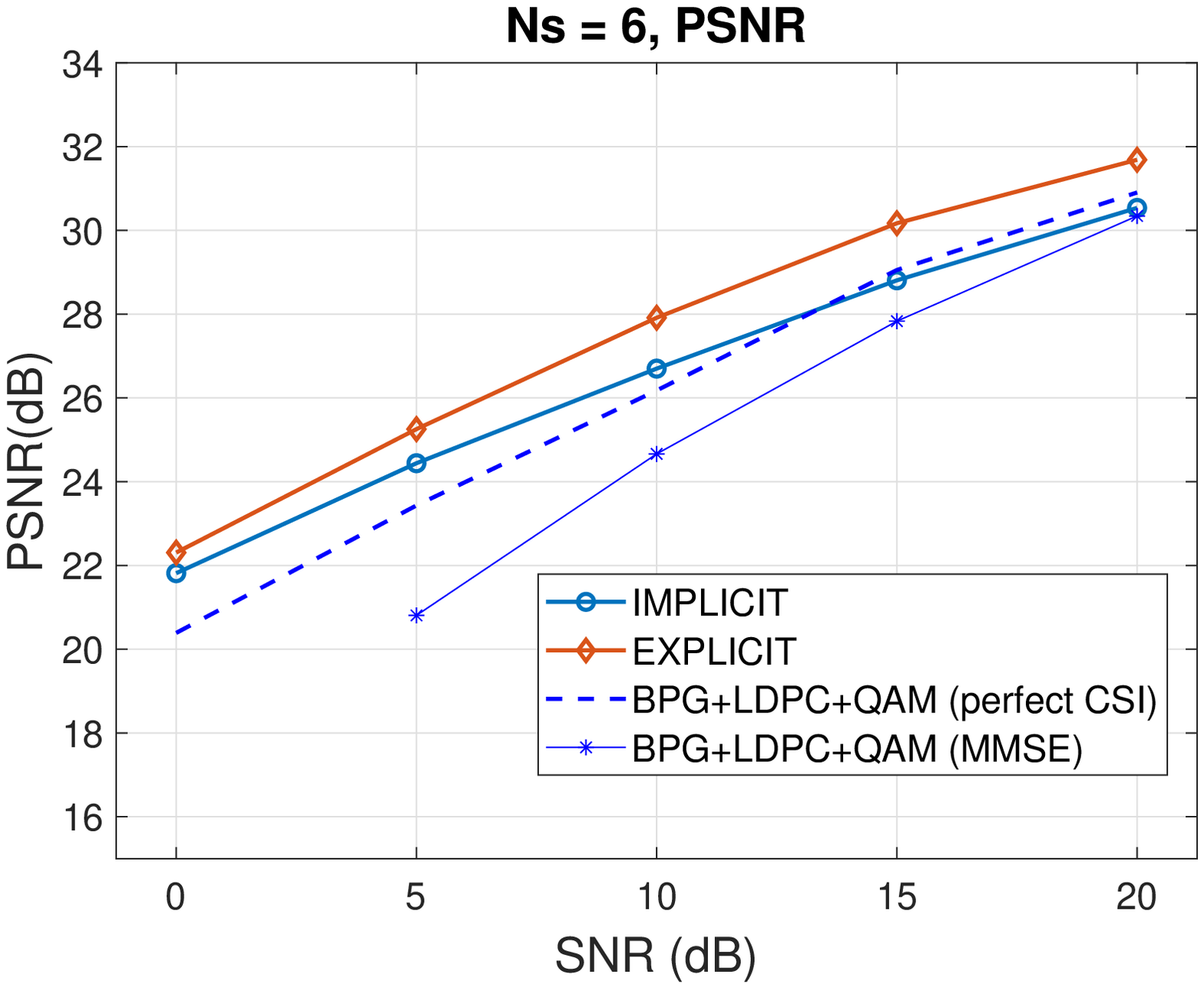}}
    \end{minipage}
  \begin{minipage}[b]{.48\linewidth}
    \centering
    \centerline{\includegraphics[width=4.7cm]{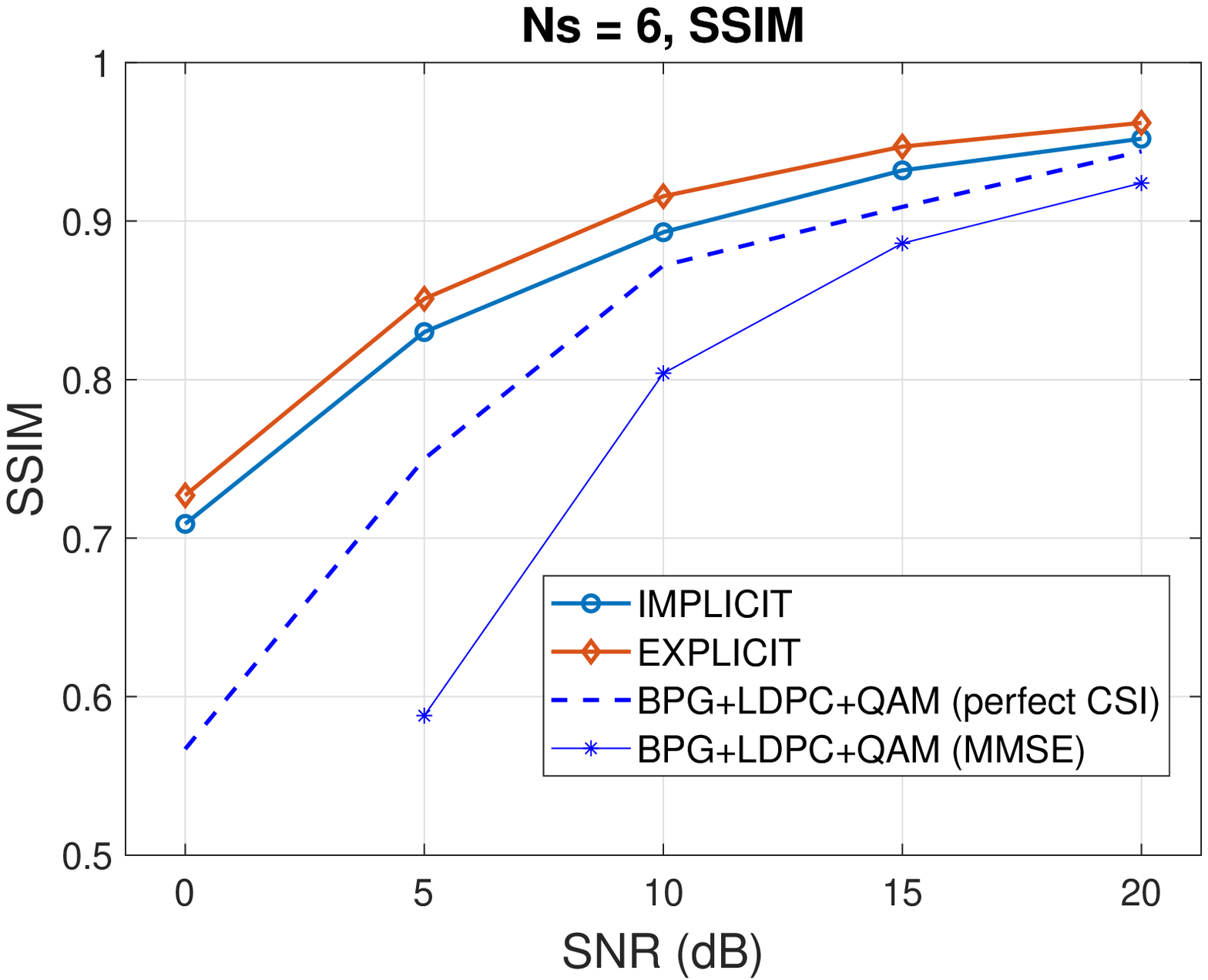}}
  \end{minipage}
  \hfill
\caption{Performance of the proposed method on CIFAR-10 with respect to SNR for a fixed number of information symbols ($N_s = 6$)}
\label{fig:fig_snrfr}
\end{figure}
}

\newcommand{\figclip}{
  \begin{figure}[t]
    \begin{minipage}[b]{.48\linewidth}
      \centering
      \centerline{\includegraphics[width=4.7cm]{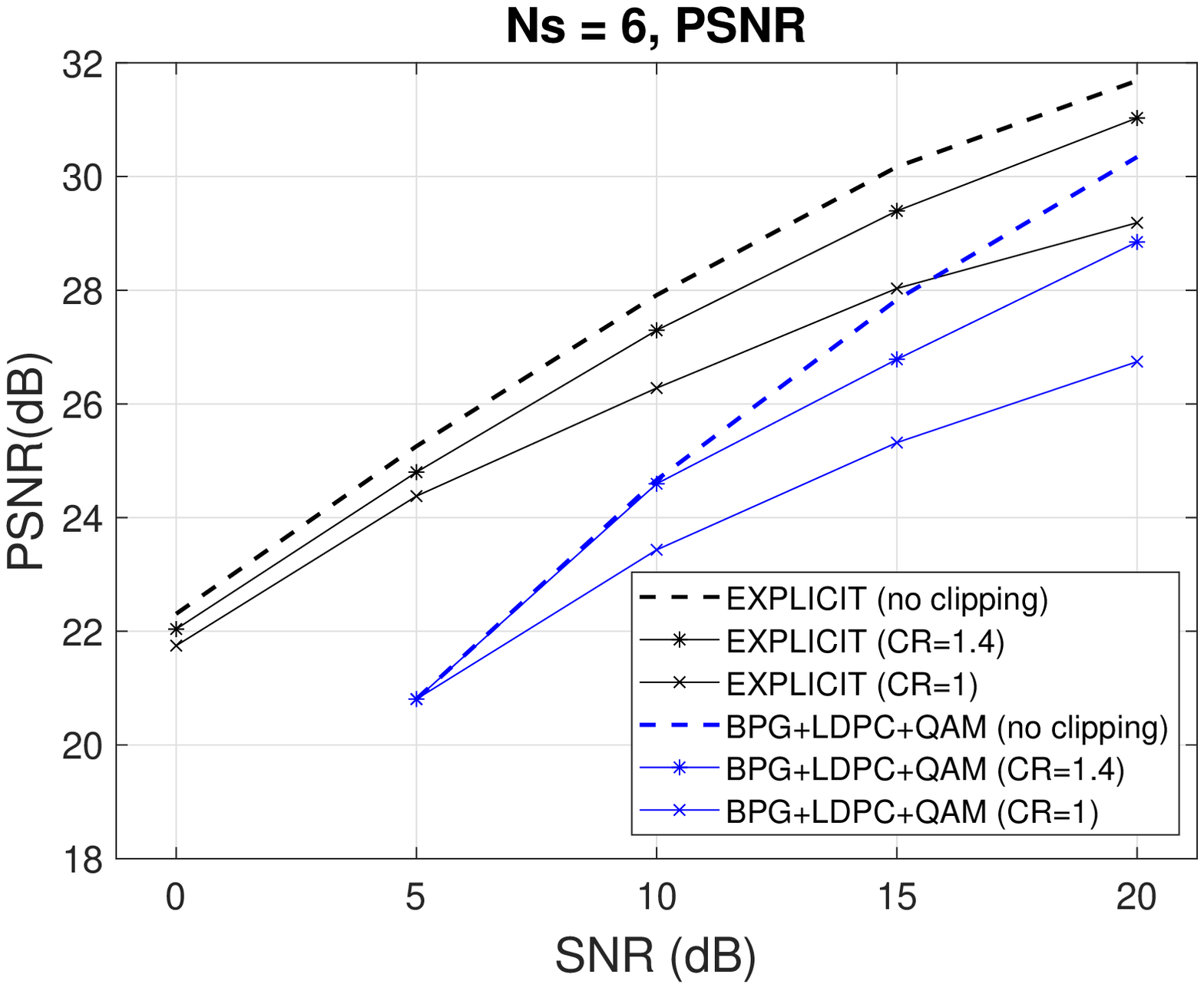}}
    \end{minipage}
  \begin{minipage}[b]{.48\linewidth}
    \centering
    \centerline{\includegraphics[width=4.7cm]{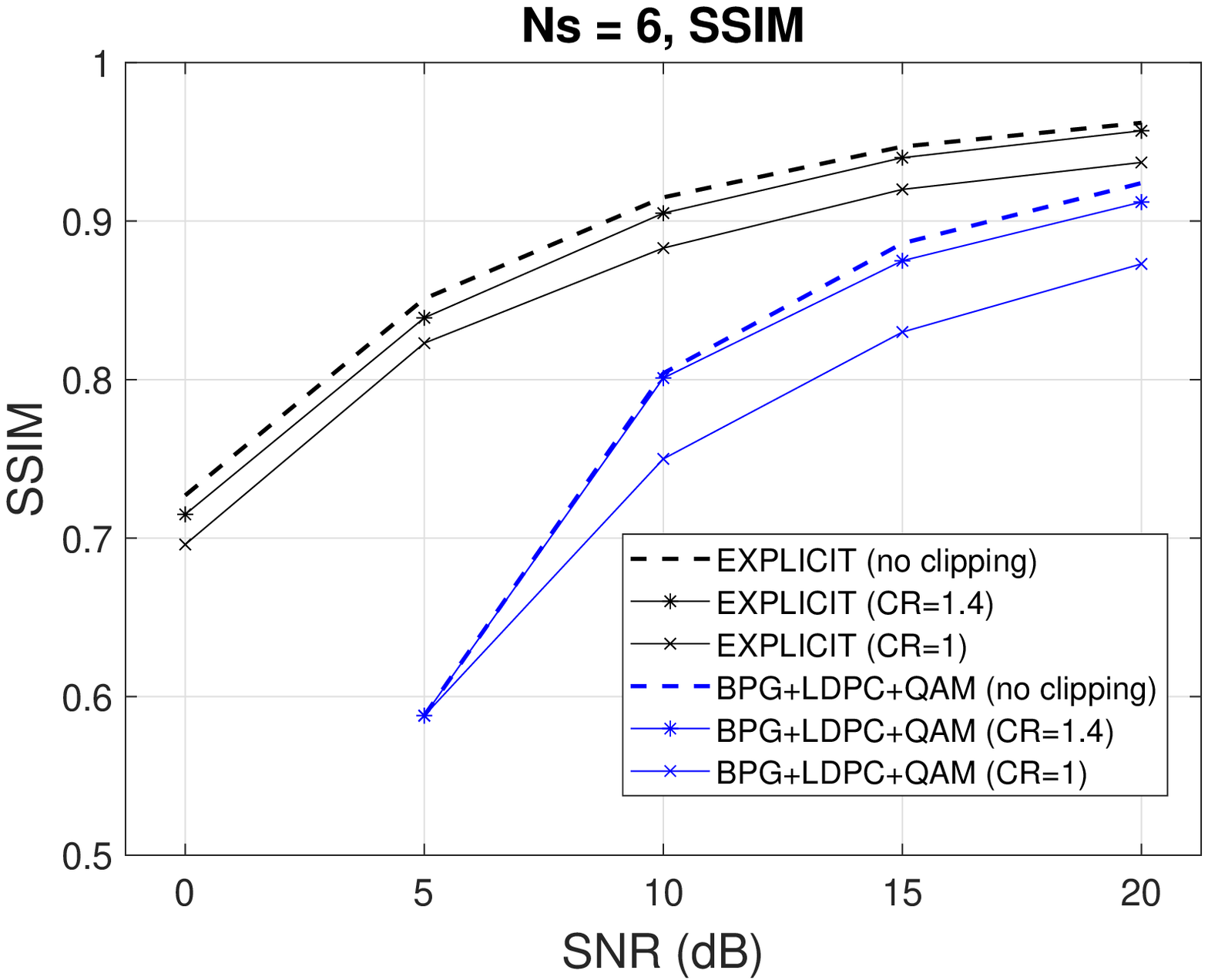}}
  \end{minipage}
  \hfill
\caption{Performance of the proposed method on CIFAR-10 with respect to SNR for different clipping ratios. $N_s = 6$.}
\label{fig:fig_clip}
\end{figure}
}

\newcommand{\figrob}{
  \begin{figure}[t]
    \begin{minipage}[b]{.48\linewidth}
      \centering
      \centerline{\includegraphics[width=4.7cm]{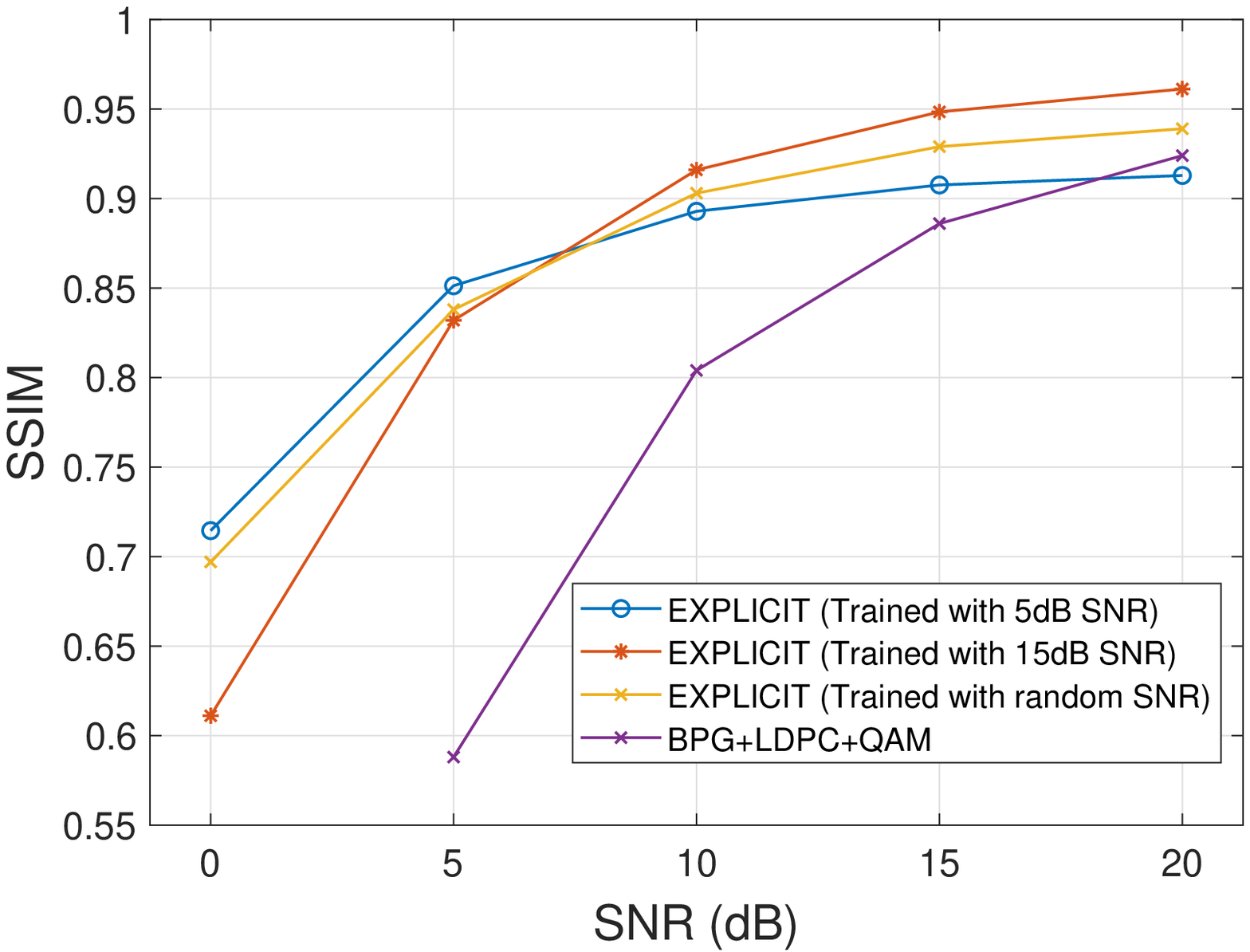}}
    \end{minipage}
  \begin{minipage}[b]{.48\linewidth}
    \centering
    \centerline{\includegraphics[width=4.7cm]{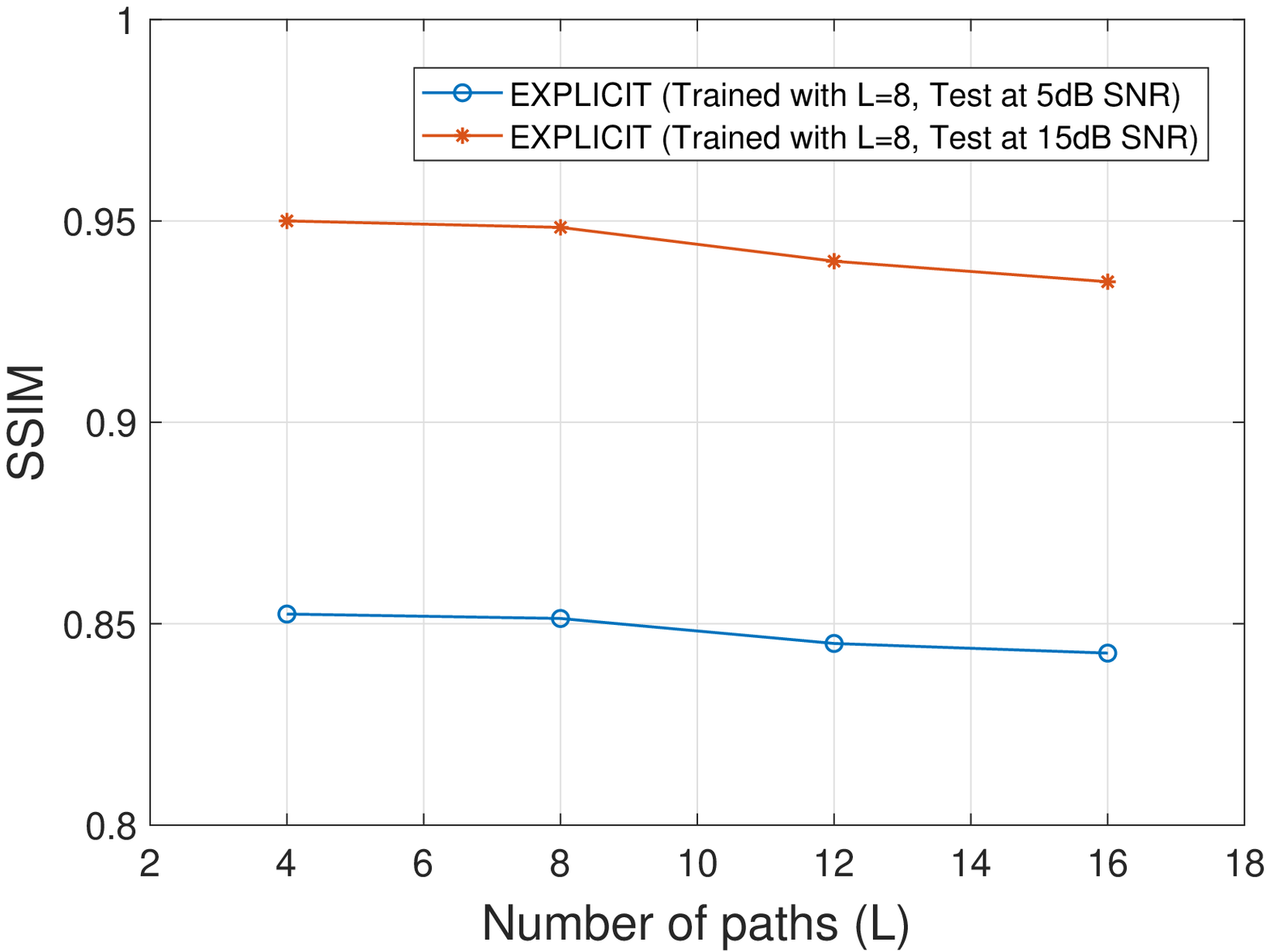}}
  \end{minipage}
  \hfill
\caption{Robustness test for the JSCC with CIFAR-10 images and $N_s=6$. Left: training vs. testing SNR mismatch scenarios, Right: multipath channel model parameter ($L$) mismatch scenarios with SNR = 5dB or 15dB.}
\label{fig:fig_rob}
\end{figure}
}

\usepackage{url}

\begin{document}

\title{Deep Joint Source Channel Coding for Wireless Image Transmission with OFDM\\
\thanks{This work was funded in part by DARPA YFA \#D18AP00076 and NSF CAREER \#1942806. }
}

\author{\IEEEauthorblockN{Mingyu Yang, Chenghong Bian, and Hun-Seok Kim}
\IEEEauthorblockA{University of Michigan, Ann Arbor, USA}
}

\maketitle

\begin{abstract}
We present a deep learning based joint source channel coding (JSCC) scheme for wireless image transmission over multipath fading channels with non-linear signal clipping. The proposed encoder and decoder use convolutional neural networks (CNN) and directly map the source images to complex-valued baseband samples for orthogonal frequency division multiplexing (OFDM) transmission. The proposed model-driven machine learning approach eliminates the need for separate source and channel coding while integrating an OFDM datapath to cope with multipath fading channels. The end-to-end JSCC communication system combines trainable CNN layers with non-trainable but differentiable layers representing the multipath channel model and OFDM signal processing blocks. Our results show that injecting domain expert knowledge by incorporating OFDM baseband processing blocks into the machine learning framework significantly enhances the overall performance compared to an unstructured CNN. Our method outperforms conventional schemes that employ state-of-the-art but separate source and channel coding such as BPG and LDPC with OFDM. Moreover, our method is shown to be robust against non-linear signal clipping in OFDM for various channel conditions that do not match the model parameter used during the training.
\end{abstract}


\section{Introduction}
Shannon's separation theorem\cite{shannon1948mathematical} states that, as the size of the transmitted message goes to infinity for memory-less channels, it is optimal to split the communication task into (i) removing the redundant information of the source as much as possible and (ii) re-introducing redundancy for message reconstruction in the presence of channel noise. Based on this famous theorem, most modern systems for wireless image transmission first compress the image with a source coding algorithm (e.g., JPEG, WebP, BPG) and then encode the bit stream with a source-independent channel code (e.g., LDPC, Polar, Turbo, etc.) as shown in Figure \ref{fig:fig_intro}. This separate source and channel coding is attractive for practical communication systems because of the convenient modularity. However, in practice, we often transmit a finite number of bits and that breaks the assumption of Shannon's separation theorem. For a finite length image source, reconstruction quality becomes a joint function of the source coding distortion and channel coding error. Thus it is difficult to design an optimal scheme that balances these two sources of the quality degradation. Besides, the optimal maximum-likelihood detection for the joint source and channel coded information is NP-hard in general. Considering the complexity and power constraints of practical systems, sub-optimal separate source and channel coding is regarded as a standard scheme.

\figintroJSCC
\figofdmlarge
In recent years, deep learning has been successfully applied to a wide range of areas such as computer vision \cite{krizhevsky2012imagenet}\cite{simonyan2014very} and natural language processing \cite{cho2014learning}\cite{vaswani2017attention}, and it achieved significant performance improvement over analytical (non-data-driven) algorithms. Its ability to extract complex features from images has led to applications of deep learning to joint source channel coding (JSCC) over simple channel models such as additive white Gaussian noise (AWGN) \cite{bourtsoulatze2019deep}\cite{kurka2019successive}\cite{kurka2020deepjscc} and binary symmetric channel (BSC) \cite{choi2019neural}\cite{song2020infomax}. These methods demonstrate the feasibility of practical machine learning based JSCC schemes that yield better performance than prior separate coding schemes. However, there is no JSCC work that extends to realistic and challenging multipath fading channels. 

In this paper, we propose a deep learning based JSCC scheme for wireless image transmission under multipath fading channels. Inspired by prior works \cite{bourtsoulatze2019deep}\cite{kurka2019successive}\cite{kurka2020deepjscc}, we employ a pair of encoder and decoder that directly encodes images as complex valued channel input samples and recovers them from noisy channel outputs. Moreover, adopting the idea of model-driven deep learning approach\cite{he2019model}, we extend the JSCC framework to combine trainable CNN layers as well as non-trainable but differentiable layers representing a multipath  channel  model  and OFDM baseband processing blocks. As the authors of other prior work reported in\cite{o2017introduction}\cite{felix2018ofdm}, incorporating the domain expert knowledge improves the training speed and enhances the performance over a conventional auto-encoder approach without an explicit domain-knowledge-driven structure. We show that as we impose additional domain-knowledge-driven structure such as channel estimation and equalization into the decoder, its performance further improves. In the proposed scheme, OFDM processing blocks (FFT, cyclic prefix insertion, etc.) and multpath channel model are both implemented as differentiable layers so that the end-to-end framework can be trained with multipath fading as well as non-linear signal clipping at the transmitter. Our results show that the proposed JSCC method outperforms the baseline separate source and channel scheme with a state-of-the-art image compression method and error correction code applied to OFDM. Furthermore, we show that our proposed method is robust for a wide range of SNR and channel conditions that the neural network has not seen during the training process.

\section{Related Work}

\subsection{Deep Joint Source Channel Coding}
Deep JSCC was first studied in \cite{bourtsoulatze2019deep} where the authors incorporate both source and channel coding into an auto-encoder structure, and jointly learn source compression and error correction coding through the back-propagation optimization. The proposed deep JSCC algorithm outperforms conventional separate schemes under AWGN and Rayleigh flat fading channel. It can be further extended with feedback \cite{kurka2020deepjscc} or progressive transmission \cite{kurka2019successive}. Similarly, deep JSCC on BSC and binary erasure channels have been studied in \cite{choi2019neural}\cite{song2020infomax}. Due to their discrete nature, the authors perform the end-to-end optimization in a variational manner and optimize a lower bound of mutual information between the source images and noisy channel outputs. In additional to images,  JSCC has been studied for Gaussian source \cite{saidutta2019joint} and text source \cite{farsad2018deep}.

\subsection{Model-driven Machine Learning for Communications}

Recent years, deep learning methods have been widely applied to wireless  communications \cite{wang2017deep}\cite{qin2019deep}\cite{gui2018deep}. However, as discussed in \cite{he2019model}, these data-driven methods simply treat standard neural networks as a black box, which require a large amount of data and long training time. To address this issue, model-driven deep learning frameworks combine  standard neural networks with analytical/mathematical models and signal processing blocks guided by the expert knowledge to shorten the training time and also to improve the performance. Combining machine learning with expert domain knowledge has shown great success in many applications such as radio transformer network (RTN) \cite{o2017introduction}, symbol modulation and demodulation for OFDM \cite{felix2018ofdm}, channel estimation \cite{gao2018comnet}, and peak-to-average power Ratio (PAPR) reduction \cite{kim2017novel}.

\section{Proposed Method}

\subsection{Deep JSCC in multipath fading channels}
The general framework for deep learning based JSCC is shown in Figure \ref{fig:fig_intro}. In our approach, a pair of encoder $E_\theta$ and decoder $D_\phi$ directly map the source input $x$ to complex-valued (modulated) baseband samples $y$ and estimate the source signal $\hat{x}$ from noisy multipath fading channel outputs $\hat{y}$. Here, $\theta$ and $\phi$ represent the parameters for the encoder and decoder neural networks respectively. The channel can be represented with the conditional probability $p(\hat{y}|y; \epsilon)$ where $\epsilon$ denotes the parameters that describe the channel statistics. The channel is defined in the continuous time domain assuming interpolation and Nyquist sampling for the channel inputs and outputs.  

We parameterize the multipath fading channel using a channel transfer function;
\begin{equation}
    \hat{y} = h(y;\sigma_0^2, \ldots, \sigma_{L-1}^2, \sigma^2) = h * y + w
    \label{eq:channel}
\end{equation}
where $*$ denotes the convolution operation, $h \in \mathbb{C}^L$ denotes the sample space channel impulse response, and $L$ is the number of multipaths. $w \sim \mathcal{CN}(0, \sigma^2I_k) $ represents the additive Gaussian noise. Each path experiences independent Rayleigh fading satisfying $h_l \sim \mathcal{CN}(0, \sigma_l^2)$ for $l=0,1,\ldots, L-1$. The power for each path follows an exponential decay profile $\sigma_l^2 = \alpha_l e^{-\frac{l}{\gamma}}$ where $\alpha_l$ is a normalization coefficient to satisfy $\sum_{l=0}^{L-1} \sigma_l^2 = 1$. $\gamma$ is the time decay (or delay spread) constant. Note that the channel transfer function is fully differentiable given realizations of random channel parameters. It means that the gradients from the decoder can propagate back to the encoder through a particular realization of the multipath fading channel when the end-to-end system is trained with gradient descent methods. For the gradient descent training of the encoder and decoder pair, we use random realizations of the channel model described in this section. 

\subsection{JSCC with OFDM Extension and Signal Clipping}

We extend the JSCC framework to an OFDM-based system to efficiently mitigate the multipath fading channel with simple single-tap frequency domain equalization. We assume that each source image $x$ is transmitted in a single OFDM packet that contains $N_p$ pilot symbols and $N_s$ information symbols. For channel estimation, we adopt block-type pilots by sending known symbols on all subcarriers. For simplicity, we assume that the transmitter and receiver are synchronized in time (within the cyclic-prefix length $L_{cp}$) without any carrier frequency offset. 

The OFDM extension to the JSCC framework is shown in Figure \ref{fig:fig_ofdm}. With OFDM, the encoder output becomes the frequency domain symbols $Y\in \mathbb{C}^{N_s \times L_{fft}}$ fed into the OFDM transmitter. The pilot symbols $Y_p \in \mathbb{C}^{N_p \times L_{fft}}$ are known to both the transmitter and receiver. Here,  $L_{fft}$ denotes the number of subcarriers in an OFDM symbol. We apply inverse discrete Fourier transform (IDFT) to each OFDM symbol and append the cyclic-prefix (CP). 

One notable drawback of OFDM is the high PAPR that causes excessive power consumption at the power amplifier \cite{li1997effects}. A number of solutions have been proposed to reduce the PAPR. Among them, signal clipping is one of the simplest and extensively studied methods \cite{ochiai2002performance}\cite{kim2011power}. Suppose we have the original TX sample $y_{preclip}[n] = A[n]e^{j\phi[n]}$ where $A[n]$ is the amplitude and $\phi[n]$ is the phase. The clipping operation can be expressed as:
\begin{equation}
    y[n] = \begin{cases}
A[n]e^{j\phi[n]} &\text{when $A[n] \leq \rho \sqrt{P_s}$}\\
\rho \sqrt{P_s}e^{j\phi[n]} &\text{otherwise}
\end{cases}
\label{eq:clip}
\end{equation}
where $\rho$ denotes the clipping ratio (CR) and $P_s$ is the input signal power. In the proposed JSCC, this clipping operation is considered as an additional non-linear activation function as shown in Fig. \ref{fig:fig_net} and the gradients propagate through this clipping layer during the JSCC encoder-decoder pair training. 

After clipping, the transmit signal $y \in \mathbb{C}^{(N_s+N_p)(L_{fft}+L_{cp})}$ propagates through the multipath channel described in (\ref{eq:channel}). Once the receiver obtains the noisy channel output $\hat{y}$, it removes the CP, and applies DFT to produce the frequency domain pilots $\hat{Y}_p$ and data symbols $\hat{Y}$. If there was no clipping (either deliberate or amplifier-inherent), the following holds for pilot and information symbols:
\begin{equation}
    \hat{Y}_p[i,k] = H[k]Y_p[i,k] + W[i,k]
    \label{eq:h1}
\end{equation}
\vspace{-0.3cm}
\begin{equation}
    \hat{Y}[j,k] = H[k]Y[j,k] + V[j,k]
    \label{eq:h2}
\end{equation}
where $H$ denotes the channel frequency response for the $k_{th}$ subcarrier, and both $W$ and $V$ denote noise samples. The decoder is trained to estimate transmitted source $\hat{x}$ given $Y_p$, $\hat{Y}_p$ and $\hat{Y}$. The loss function of this learning problem can then be expressed as:
\begin{equation}
    L(\theta, \phi) = \mathbb{E}_{p(x, \hat{Y}, \hat{Y}_p)}[d(D_\phi(E_\theta(x),\hat{Y},\hat{Y}_p), x)]
    \label{eq:loss}
\end{equation}
where $\mathbb{E}_p[\ ]$ is the expected value over distribution $p$, and $d$ denotes the distortion calculated by $x$ and $\hat{x} (= D_\phi(E_\theta(x),\hat{Y},\hat{Y}_p))$. The goal is to find $\theta^*$ and $\phi^*$ that minimize the loss function $L(\theta, \phi)$. 

It must be noted that including OFDM signal processing into the proposed JSSC framework is possible because IDFT/DFT and CP removal/insertion can be treated as linear layers in the neural network with fixed parameters. During back-propagation, gradients can pass through these (I)DFT/CP linear layers. Similarly, the proposed JSSC framework learns the encoder and decoder networks in the presence of signal clipping at the transmitter which is similar to a clipped ReLu activation function.

\subsection{Decoder Design with Domain Knowledge}

To estimate the source information, one possible approach is to concatenate $\hat{Y}$, $Y_p$ and $\hat{Y}_p$ to directly feed them into a deep (`upscale') neural network. This is illustrated in Figure \ref{fig:fig_ofdm} as the \textit{IMPLICIT} method. This approach relies on the deep neural network to learn the underlying relationships in (\ref{eq:h1}) and (\ref{eq:h2}) to perform the (sub)optimal estimation of the source. We name this method \textit{IMPLICIT} because it implicitly learns traditional channel estimation and equalization processes in the decoder network. However, this method treats the entire decoder as a blockbox, and hence it faces the problems of slow convergence and sub-optimal performance as argued in \cite{he2019model}. As an alternative solution, we propose a decoder structure shown in Figure \ref{fig:fig_ofdm} bottom right where explicit signal processing kernels such as channel estimation and equalization are included as pre-processing steps in front of the neural network based decoder. Due to the signal clipping and other non-linear activations in the transmitter, the manually-designed channel estimation and equalization methods may not be optimal. Thus, we further introduce two residual connections with light-weight neural networks (`subnets') so that they can learn and compensate residual errors in the channel estimation and equalization. We name this method \textit{EXPLICIT}. For simplicity, we use the per-channel MMSE channel estimation method \cite{felix2018ofdm} which does not need second order statistics of the channel and matrix inversion. For equalization, we adopt a conventional MMSE equalizer.

\section{Training and Evaluation}

\fignet

We test our proposed method using CIFAR-10 and CelebA datasets. The CIFAR-10 dataset contains 60,000 32$\times$32-pixel images whereas CelebA contains more than 200,000 celebrity images. We scaled and cropped CelebA images to 64$\times$64-pixels. For testing, we take 10,000 images from each test dataset (unused for training) and transmit each image 5 times through different random realizations of the multipath channel. We use both PSNR and SSIM to evaluate the reconstruction quality. We test our JSCC method on CIFAR-10 using the network structure shown in Figure \ref{fig:fig_net}, which follows the design principles in \cite{isola2017image}. Because of the image size difference, networks for CelebA have one more down-sampling module at the encoder and one more upsampling module at the decoder. The subnets in \textit{EXPLICIT} method follow a \textit{Conv-BatchNorm-ReLu-Conv-BatchNorm} structure just like the residual block in the main network shown in Figure \ref{fig:fig_net}. 

We adopt mean-squared-error (MSE) as our distortion measure (\ref{eq:loss}) for the network training. For both datasets, we use a batch size of 128 and train the neural networks using ADAM with $\beta_1=0.5$ and $\beta_2 = 0.999$. For CIFAR-10, we apply a learning rate of $10^{-3}$ and train for 400 epochs with linear learning rate decaying for the last 200 epochs. For CelebA, the learning rate is set to $5\times10^{-4}$ and we train the end-to-end system for 60 epochs with linear learning rate decaying for the last 30 epochs. 

The OFDM and channel parameters are set to $L_{fft}=64, L_{cp}=16, N_p=2, L=8$, and $\gamma=4$. Note that the coding rate of our method depends on the number of  samples after encoding. For a source image of the size $H\times W \times C$ pixels ($C$ is the number of color channels), the coding rate is $\frac{(N_p+N_s)(L_{fft}+L_{cp})}{HWC}$ (channel-use per pixel, CPP). The length of CP has to be larger than the delay spread of the channel to avoid inter-symbol interference. 


\begin{table}[tbp]
\caption{CIFAR-10 and CelebA evaluation (Top: PSNR, Bottom: SSIM)}
\begin{center}
\begin{tabular}{c|ccc|ccc}
\hline
\textbf{Dataset}&\multicolumn{3}{|c|}{\textbf{CIFAR-10, CPP=0.21}} &\multicolumn{3}{|c}{\textbf{CelebA, CPP=0.05}} \\
\cline{1-7} 
\textbf{SNR} & \textbf{0dB}& \textbf{10dB}& \textbf{20dB} & \textbf{0dB}& \textbf{10dB}& \textbf{20dB}\\
\hline
Direct w/o OFDM & 21.05& 24.93& 26.21& 21.86 & 24.76& 25.73\\
\textit{IMPLICIT} & 21.82& 26.7& 30.54& 22.27& 26.54& 29.37\\
\textit{EXPLICIT} & \textbf{22.31}& \textbf{27.91}& \textbf{31.69}& \textbf{22.92} & \textbf{27.27}&\textbf{30.31} \\
\hline
\end{tabular}
\label{tab2}
\end{center}

\begin{center}
\begin{tabular}{c|ccc|ccc}
\hline
\textbf{Dataset}&\multicolumn{3}{|c|}{\textbf{CIFAR-10, CPP=0.21}} &\multicolumn{3}{|c}{\textbf{CelebA, CPP=0.05}} \\
\cline{1-7} 
\textbf{SNR} & \textbf{0dB}& \textbf{10dB}& \textbf{20dB} & \textbf{0dB}& \textbf{10dB}& \textbf{20dB}\\
\hline
Direct w/o OFDM & 0.697& 0.84& 0.871& 0.683 & 0.788& 0.811\\
\textit{IMPLICIT} & 0.709& 0.893& 0.952& 0.702& 0.85& 0.908\\
\textit{EXPLICIT} & \textbf{0.727}& \textbf{0.915}& \textbf{0.962}& \textbf{0.73} & \textbf{0.867}&\textbf{0.924} \\
\hline
\end{tabular}
\label{tab3}
\end{center}
\end{table}

\subsection{Effect of applying domain knowledge}
\label{ddd}

To quantify the effect of incorporating domain knowledge into our framework, we compare our method with a neural network-only scheme which we call `direct' where symbols are transmitted through the multipath channel without OFDM layers. All compared methods are tested with both CIFAR10 and CelebA under different SNRs and the coding rate is kept the same for each dataset (CPP is 0.21 and 0.05 for CIFAR-10 and CelebA, respectively). The results are shown in Table \ref{tab2}. Including OFDM (I)DFT/CP processing layers, our \textit{IMPLICIT} and \textit{EXPLICIT} method attain superior PSNR and SSIM than the direct method without OFDM layers especially for high SNRs. As we introduce more domain knowledge into \textit{EXPLICIT} configuration, the system experiences additional performance gain. Note that the subnets in \textit{EXPLICIT} (Fig. \ref{fig:fig_ofdm}) only introduce 0.2\% additional parameters than IMPLICIT method to obtain the gain shown in Table \ref{tab2}.  

\figlr

As shown in Figure \ref{fig:fig_lrcurve}, significant training efficiency gain is obversed for \textit{IMPLICIT} and \textit{EXPLICIT} compared to the `direct' method. With OFDM (I)DFT/CP layers, \textit{IMPLICIT} and \textit{EXPLICIT} attain significantly (2.5$\times$) lower loss function values with faster ($5\times$) training convergence time for CelebA dataset. The training time is similar for \textit{IMPLICIT} and \textit{EXPLICIT} schemes but \textit{EXPLICIT} method converges to a 20\% lower loss function value for CelebA training.

\subsection{Comparison with a separate coding scheme}

In this section, we compare our proposed JSCC method with a baseline separate source and channel coding scheme. The baseline separate coding scheme uses the state-of-the-art BPG\cite{BPG} as the image codec and LDPC in the IEEE 802.11n WiFi standard \cite{LDPC} as the channel code. We test with three different LDPC codes $(972, 1944), (1296, 1944), (1458, 1944)$ which correspond to rates $1/2$, $2/3$ and $3/4$ respectively. We use BPSK, QPSK, 16QAM, and 64QAM for OFDM modulation applied to BPG and LDPC coded bit sequences. Then, we enumerated all possible modulation and coding rate combinations to identify the optimal configuration for the baseline separate coding scheme. The channel estimation for this baseline is based on per-channel MMSE, which is also used in our \textit{EXPLICIT} scheme. The estimated channel is then used to calculate the log likelihood ratio (LLR) for LDPC decoding in the baseline separate coding scheme. 

We compare PSNR and SSIM of the proposed JSSC and the separate coding baseline at the same/similar rate and SNR. Because BPG cannot achieve arbitrary compression rates, we exhaustively search for the BPG parameter to obtain a rate closest to the target. For fair comparison, we did not include the BPG headers in the BPG rate computation. When BPG-LDPC encoded packets fail to reconstruct images due to fatal bit errors, we simply assume they achieve the average PSNR/SSIM of decodable images, giving an advantage to the baseline separate coding scheme.  

Figure \ref{fig:fig_lowsnr} and \ref{fig:fig_highsnr} compare the performance of 
deep JSCC schemes and baseline approaches with respect to the number of information symbols $N_s$ in different SNR regimes using CIFAR-10 dataset. Smaller $N_s$ indicates more aggressive (lower) rate to send an entire image with a fewer number of data symbols. For comparison, we also show the performance of baseline separate coding schemes with perfect channel state information (CSI) for error-free channel estimation. Note that, for the $\leq$5dB SNR case, BPG with 1/2 LDPC+BPSK cannot achieve the rate of $N_s=4$ for the full image. Figure \ref{fig:fig_lowsnr} and \ref{fig:fig_highsnr} show our deep JSSC consistently outperforms baseline separate coding schemes  especially for the low SNR and low rate regime. Note that the performance of the baseline method approaches to that of the deep JSSC for some high rates when perfect CSI is given for their advantage. For the deep JSSC (both \textit{IMPLICIT} and \textit{EXPLICIT}), we never provide perfect CSI to the decoder. With imperfect CSI (same as in JSSC), the perfomrnace of the baseline scheme is significantly worse (e.g., $>$3dB PSNR) at the same low rate target. It is known that SSIM is a better metric than PSNR to quantify human perception of an image. Our deep JSSC provides a significant SSIM gain compared to the baseline although it is trained to minimizes the MSE loss function, not SSIM. 

\figlowpsnr
\fighighpsnr

For the evaluation in Figure \ref{fig:fig_snrfr}, we fix the rate to $N_s=6$ for the the CIFAR-10 dataset (0.21 channel-use per pixel) and evaluate the reconstructed image quality for a wider SNR range. For the baseline method, we choose at each SNR point the optimal combination of the OFDM modulation QAM size and LDPC rate that yields the best performance. For 0dB SNR, the baseline method without perfect CSI failed to provide reasonable PSNR/SSIM at the given rate. It is observed that the proposed \textit{EXPLICIT} JSCC method consistently outperforms the baseline with $>$3dB SNR gain for the same PSNR. The SNR gain is even higher when the image quality is measured in SSIM. 
\figpsnrfr




\subsection{Robustness Analysis}

Finally, we evaluate robustness of the JSCC with respect to the signal clipping. We also evaluate the impact of mismatched channel statistics such as SNR or the number of multipaths used during the JSCC training vs. actual testing.

\figclip

Figure \ref{fig:fig_clip} shows the effect of clipping to the performance of the \textit{EXPLICIT} JSCC and baseline method with various clipping ratios (CR) of $\rho=1$, $1.4$, and $\infty$ as defined in (\ref{eq:clip}). Note that for the baseline, we adopt the methods in \cite{ochiai2002performance}\cite{kim2011power} to calculate the LLR including the non-linear distortion introduced by clipping. It is  observed that our JSCC provides a graceful degradation in PSNR/SSIM than the baseline method as more aggressive clipping is applied. And same as the previous section, our method outperforms the baseline in all range of SNRs in the presence of clipping. 

 Figure \ref{fig:fig_rob} left shows the JSCC performance when it is trained for a specific SNR (5dB or 15dB) and tested with a wide range of SNR from 0dB to 20dB. It is observed that the model trained with 5dB performs better for the low SNR regime while the model trained with 15dB performs better in the high SNR region. More balanced JSCC performance is obtained when the model is trained with random SNR values covering a wide range. The \textit{EXPLICIT} JSCC model in Figure \ref{fig:fig_rob} right was trained with an $L=8$ multipath channel model and tested with a wide range of $L$ but the same noise level. Two curves show the performance at 5dB and 15dB SNR cases. It is confirmed that our JSCC method is relatively insensitive to the mismatch in the number of multipaths of the channel model for the training vs. testing.

\figrob

\section{Conclusion}

In this paper, we present a deep learning based JSCC scheme for wireless image transmission over multipath fading channels. We extend the JSCC scheme with explicit OFDM layers and signal processing steps driven by the expert domain knowledge. With addition explicit steps, the JSCC training convergence time improves with enhanced performance compared to naive \textit{IMPLICIT} approaches without exploiting domain knowledge. Through extensive experimental simulations, we show that the deep JSCC method outperforms the conventional separate source and channel coding schemes with OFDM especially for low SNR and low rate regimes. The proposed JSCC framework effectively learns source compression as well as protection against noise in challenging multipath channels. Moreover, the JSCC framework incorporates deliberate signal clipping during the training process to significantly reduce the PAPR with graceful performance degradation. Our approach is shown to be robust when the evaluation channel model characteristics are mismatched to those used during the training.       

\bibliographystyle{IEEEbib}
\bibliography{strings,refs}

\end{document}